\documentclass{ieeeaccess11}
\usepackage{cite}
\usepackage{flushend}
\usepackage{amsmath,amssymb,amsfonts}
\usepackage{algorithmic}
\usepackage{graphicx}
\usepackage{textcomp}
\usepackage[utf8]{inputenc}
\usepackage{amsfonts}
\usepackage{amssymb}
\usepackage{physics}
\usepackage{qcircuit}
\usepackage{caption}
\usepackage{graphicx}
\usepackage{multirow}
\usepackage{siunitx}
\usepackage{url}
\usepackage{float}  
\usepackage{subfigure}  
 
\usepackage{amsfonts}
\usepackage{float}
\usepackage[misc]{ifsym} 
\usepackage{hyperref}
\usepackage{algorithm}
\usepackage{algorithmic}

\usepackage{etoolbox}
\def\BibTeX{{\rm B\kern-.05em{\sc i\kern-.025em b}\kern-.08em
    T\kern-.1667em\lower.7ex\hbox{E}\kern-.125emX}}
\begin{document}
\doi{}
\title{A Low Complexity Quantum Principal Component Analysis Algorithm} 
\author{\uppercase{Chen He}\authorrefmark{1,2}, \uppercase{Jiazhen Li}\authorrefmark{1},
\uppercase{Weiqi Liu}\authorrefmark{1},
\uppercase{Z. Jane Wang}\authorrefmark{2},~Fellow,~IEEE
}
\address[1]{Northwest University, Xi'an, China}

\address[2]{The University of British Columbia, Vancouver, BC, Canada}


\begin{abstract}
In this paper, we propose a low complexity quantum principal component analysis (qPCA) algorithm. Similar to the state-of-the-art qPCA, it achieves dimension reduction by  extracting principal components of the data matrix, rather than all components of the data matrix, to quantum registers, so that samples of measurement required can be reduced considerably. However, the major advantage of our qPCA over the state-of-the-art qPCA is that it requires much less quantum gates. In addition, it is more accurate due to the simplification of the quantum circuit. We implement the proposed qPCA on the IBM quantum computing platform, and the experimental results are consistent with our expectations.
\end{abstract}

\titlepgskip=-15pt

\maketitle
\section{Introduction}
Principal component analysis (PCA) \cite{wall2003singular,karamizadeh2013overview,shlens2014tutorial,bro2014principal} is widely employed in signal processing and machine learning for dimension reduction, and has a time complexity of $O(N^3)$, where $N$ is the dimension of the data. When the dimension of the data is large, the classical PCA becomes non-tractable.  Quantum principal component analysis algorithm (qPCA) can reduce the time complexity to $O(N \text{ploy}(\log N))$\cite{lloyd2014quantum,shao2019improved,yu2019quantum} because of the quantum computer’s parallelism \cite{nielsen2002quantum}. 

\begin{figure*}
	\centering
		\includegraphics[scale=.75]{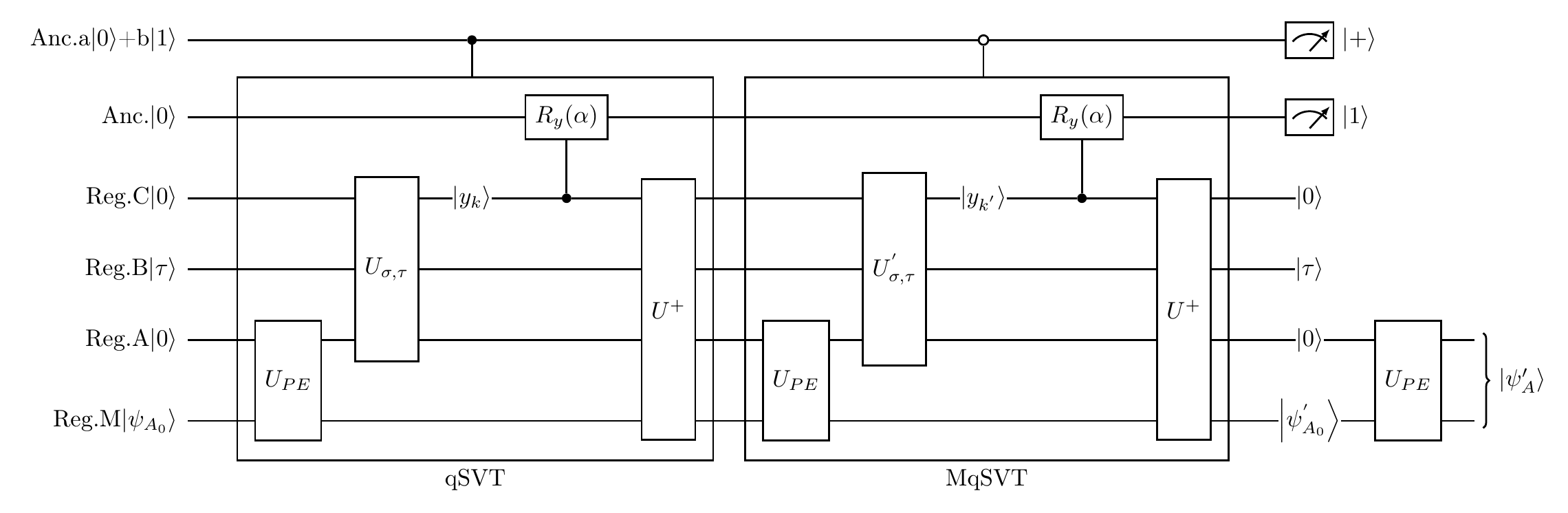}
    \caption
{\label{Fig ref qPCA circuit}The quantum circuit of the qPCA proposed in \cite{lin2019improved}, which consists of two parts:  quantum singular values threshold (qSVT) \cite{duan2018efficient} and modified qSVT (MqSVT). The input of the circuit is $\ket{\psi_{A_0}}$, and the output is $\ket{\psi_A^{'}}$, where the principal components are in the quantum registers. In the circuit $y_k=(1-\frac{\tau}{\sigma_{k}})_{+}$, $y_{k}^{'}=(1+\frac{\tau}{\sigma_{k}})_{+}$, $\tau$ is the threshold to filter out the small $\sigma_k$'s, and $\alpha$ is the parameter of rotation operation $R_y(\alpha)$, which can be adjusted to improve the success probability and fidelity of the algorithm.}
\end{figure*}

The qPCA in \cite{lloyd2014quantum} outputs the quantum state containing all the eigenvalues and eigenvectors of the data, and the top-t components (principal components) are obtained by sampling. For instance, for a matrix $A_0\in\mathbb{C}^{p\times q}$, let $A=\sum_{k=1}^{r}\lambda_k u_k u_k^{T}$ be the eigendecomposition of $A=A_0A_0^{+}$,
where $A_0^{+}$ is the conjugate transpose of $A_0$. The qPCA in \cite{lloyd2014quantum} showed that phase estimation can be employed to extract all eigenvalues $\lambda_k$ and eigenvectors $u_k$ into quantum registers with time complexity $O(r \text{ploy}(\log p))$, i.e. the qPCA outputs the quantum state
\begin{align}\label{qPCA}
\ket{\psi_{A}}=\sum_{k=1}^{r}\lambda_k\ket{\lambda_k}\ket{u_k}.
\end{align}
However, since $\ket{\psi_{A}}$ contains all the $r$ components of $A$, the qPCA in \cite{lloyd2014quantum} may need a lot of samples to obtain the principle components. To avoid this disadvantage, \cite{lin2019improved} proposed an improved qPCA as shown in Fig.~\ref{Fig ref qPCA circuit}, which yields a quantum state  containing the approximation of the components with the top $t$ $(t\ll r)$ largest eigenvalues only:
\begin{align}\label{psi_A_1}
\ket{\psi_{A}^{'}}\approx\sum_{k=1}^{t}\sigma_k\ket{\lambda_k}\ket{u_k}\ket{v_k},
\end{align}
where $\sigma_k$ are the singular values of $A_0$, $u_k$, $v_k$ are the left and right singular vectors respectively. As a result, the successful probability of obtaining a principal component increases to $\sum_{k=1}^{r}\lambda_k^{2}/\sum_{k=1}^{t}\lambda_k^{2}$ times for each measurement, and the time complexity is also reduced to $O(t \text{ploy}(\log p))$. The quantum circuit for obtaining $\ket{\psi_{A}^{'}}$ by the qPCA \cite{lin2019improved} is shown in Fig.~\ref{Fig ref qPCA circuit}.
 
\begin{figure*}
\centering
\includegraphics[scale=.8]{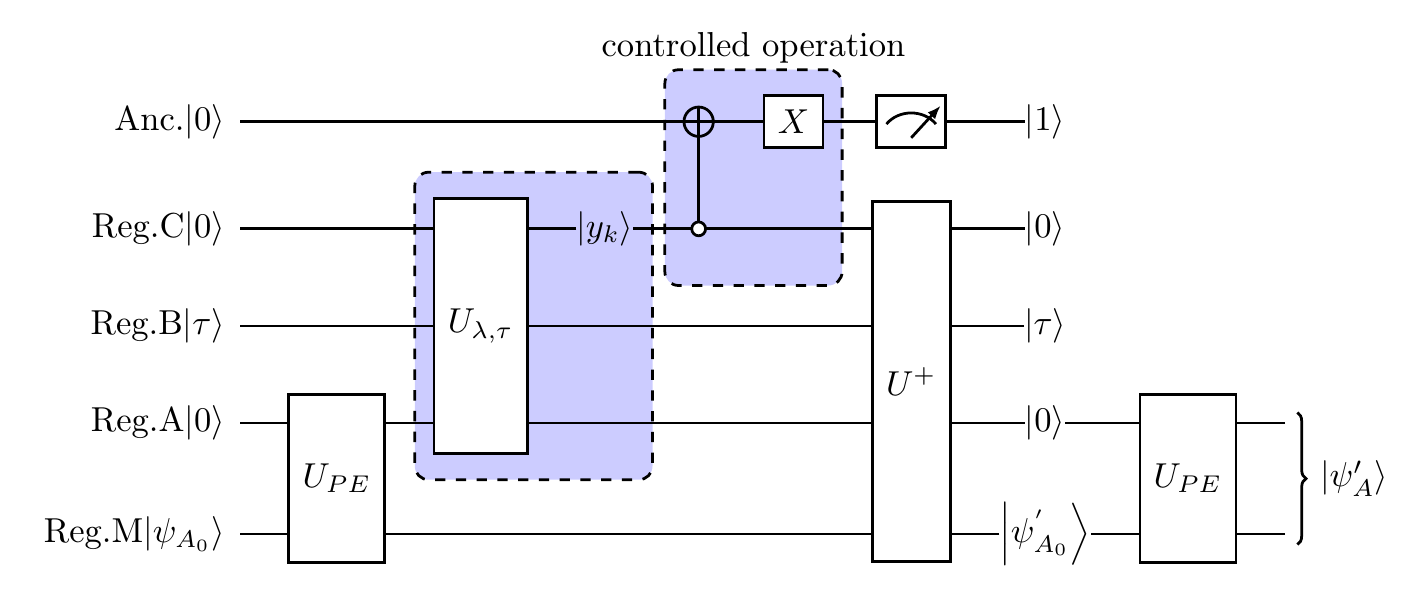}
\caption{\label{Fig our qPCA}The quantum circuit of the proposed qPCA. The input of the circuit is $\ket{\psi_{A_0}}$ and the output is the quantum state $\ket{\psi_{A}^{'}}$, where the principal components are in the quantum registers. In this circuit, $y_k=(1-\frac{\tau}{\lambda_k})_{+}$, and $\tau$ is the threshold to filter out small $\lambda_k$'s. As we can see the quantum circuit of the proposed qPCA requires much less quantum gates compared with the state-of-arts in \cite{lin2019improved}.} 
\end{figure*}

However, there are still two concerns for state-of-the-art qPCA in \cite{lin2019improved}. One is that the algorithm requires a lot of quantum gates, and the other is that the approximation is taken in two places, which may lead to a lower accuracy.
In this paper, we propose a low complexity qPCA algorithm shown in Fig.~\ref{Fig our qPCA}. Compared with the state-of-the-art qPCA \cite{lin2019improved}, the quantum circuit of our algorithm requires much less quantum gates, and the approximation is only taken in one place. The paper is organized as follows: In Section~\ref{sec2 algorithm}, we proposed a low complexity qPCA algorithm. In Section~\ref{sec3 analysis}, we analyze the complexity and accuracy of our qPCA algorithm compared with  state-of-the-art algorithm. In Section~\ref{sec4 experiement}, we implement the proposed algorithm on IBM Quantum Experience, and verify the proposed algorithm. Finally we conclude this work in Section~\ref{sec5 conclusion}. 

\begin{figure*}
\centering
\subfigure[The quantum circuit of the unitary operation $U_{\lambda,\tau}$.]{
\begin{minipage}{1.0\textwidth}
\includegraphics[scale=.85]{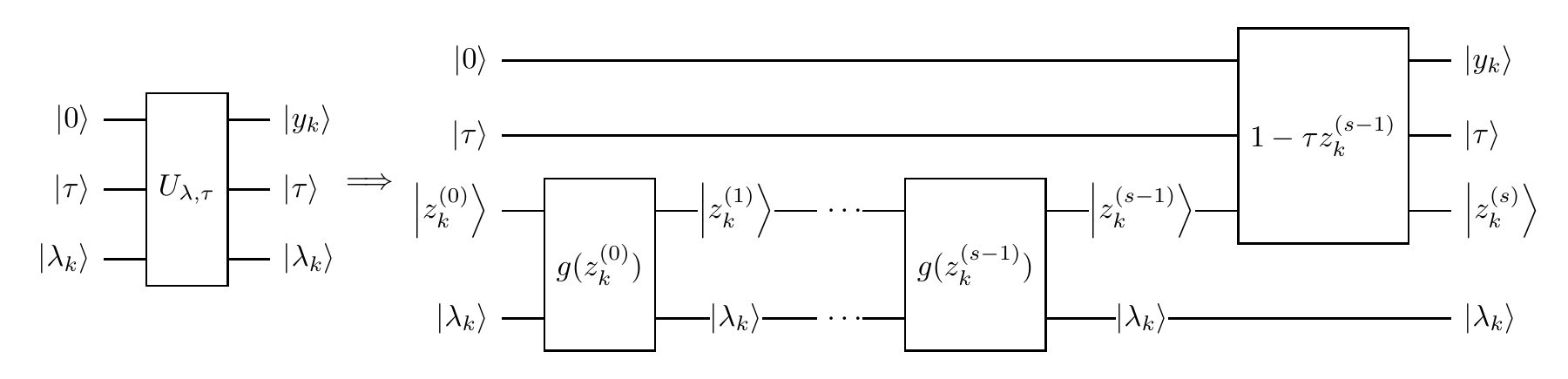}\label{Fig U_lambdatau_1}
\end{minipage}
}
\subfigure[The quantum circuit of one Newton's iteration \cite{bhaskar2015quantum} for computing $z_k^{(i+1)}$, which contains five QFT additions or QFT multiplications.]{
\begin{minipage}{1.0\textwidth}
\includegraphics[scale=.55]{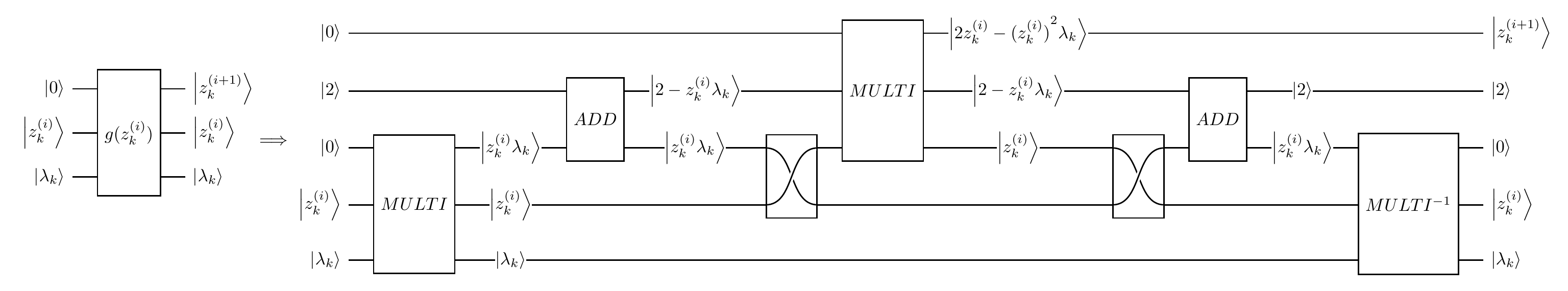}\label{Fig g(z_k)}
\end{minipage}
}

\subfigure[The quantum circuit for computing $y_k=1-\tau z_k$, which contains three QFT additions or QFT multiplications.]{
\begin{minipage}{1.0\textwidth}
\centering
\includegraphics[scale=0.65]{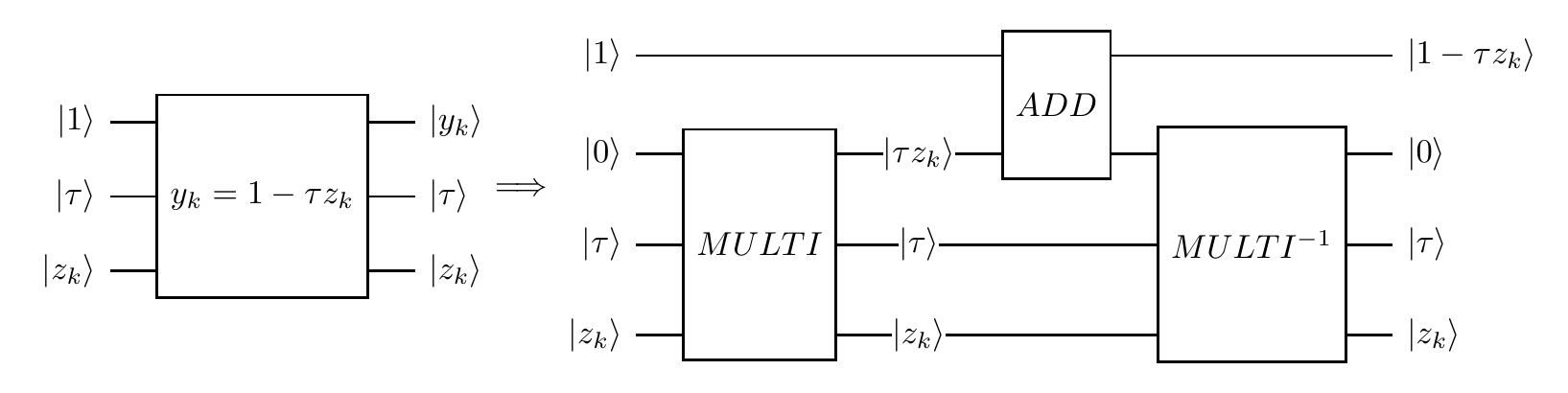}\label{Fig y_k}
\end{minipage}
}
\caption{\label{Fig U_lambdatau_}The quantum circuit of the Unitary operation $U_{\lambda,\tau}$, which contains eight QFT arithmetic operations\cite{csahin2020quantum}.} 
\end{figure*}

\section{The proposed qPCA}\label{sec2 algorithm}

In this section, we introduce the procedures of our low complexity qPCA algorithm and design the corresponding quantum circuit.

The quantum state of the matrix $A_0=\sum_{k=1}^{r}{\sigma_k}{u_k}{v_k^{T}}$ is given by \cite{duan2018efficient}
\begin{eqnarray}\label{psi_A_0}
\ket{\psi_{A_0}}=\sum_{k=1}^{r}\sigma_{k}\ket{u_k}\ket{v_k}.
\end{eqnarray}
The purpose of qPCA is to extract the larger eigenvalues of the $A_0A_0^+$ from the amplitudes to the quantum register. For instance, the qPCA in \cite{lin2019improved} extracts $\lambda_k$ from (3) to the quantum register in (2).
However, the algorithm in \cite{lin2019improved} requires a lot of quantum gates and involves approximation in two places.

In this paper we show that 
\begin{align}\label{our qPCA steps}
\ket{\psi_{A_0}}\xrightarrow{(I\otimes U_{PE})(I\otimes U^{\dagger})(CU\otimes I)(I\otimes U_{\lambda,\tau}\otimes I)(I\otimes U_{PE})}
\ket{\psi_{A}^{'}},
\end{align} 
which corresponds to a low complexity qPCA algorithm requiring less quantum gates and involving approximation in only one place. The operations in \eqref{our qPCA steps} can be decomposed into following building blocks.

\subsubsection{Phase estimation \texorpdfstring{$I\otimes U_{PE}$}{Lg}} The purpose of the phase estimation $U_{PE}$ is to extract eigenvalues to the quantum register. Suppose a unitary operator $U$ has an eigenvector $\ket{u}$ with eigenvalue $e^{2\pi i \phi}$, where the $\phi$ is unknown \cite{nielsen2002quantum}.  Phase estimation \cite{van2007optimal} can extract the phase $\phi$ into a quantum register:
\begin{align}
\ket{0}\ket{u}\xrightarrow{U_{PE}}\ket{\phi}\ket{u}.
\end{align} 
In our algorithm, the unitary operator $e^{2\pi iA}$ has the eigenvecotrs $u_k$ with eigenvalues $e^{2\pi i \lambda_k}$, and $\lambda_k$ can be estimated as 
\begin{align}
\ket{0}\ket{\psi_{A_0}}\xrightarrow{U_{PE}(A)}\sum_{k=1}^{r}\sigma_{k}\ket{\lambda_k}\ket{u_k}\ket{v_k},
\end{align} 
where
\begin{align}
U_{PE}(A)=(QFT^{\dag}\otimes I)(e^{2\pi iA}\otimes I)(H\otimes I).    
\end{align}
In addition, the register stored $\ket{\psi_{A_0}}$ requires $m = O(\log(pq))$ qubits and the register stored $\ket{\lambda_k}$ requires $n = O(\log(\kappa))$ qubits, where $\kappa$ is the condition number of the matrix $A$ \cite{Goodfellow-et-al-2016}. The number of the quantum gates required by the $U_{PE}$ operation is $O(n^2)$ \cite{nielsen2002quantum}.

\subsubsection{Unitary operation \texorpdfstring{$I\otimes U_{\lambda,\tau}\otimes I$}{Lg}} The purpose of the unitary operation $U_{\lambda,\tau}$ is to filter out the small eigenvalues from the quantum register. This is achieved by converting $\lambda_k$ to $y_k=(1-\frac{\tau}{\lambda_k})_{+}= \max\{1-\frac{\tau}{\lambda_k},0\}$. We first set the intermediate variable $z_k=\frac{1}{\lambda_k}$, which can be obtained numerically by Newton's iteration $z_k^{(i+1)}=2z_k^{(i)}-(z_k^{(i)})^{2}\lambda_k$. This iteration can be implemented by the quantum circuit Fig.~\ref{Fig g(z_k)}.  After obtaining $z_k$, $y_k=(1-\tau z_k)_+$ can be obtained by the QFT arithmetic \cite{ruiz2017quantum}, which shown in Fig.~\ref{Fig y_k}. In summary, the unitary operation $U_{\lambda,\tau}$ for our low complexity qPCA algorithm can be represented as:
\begin{align}
\ket{0}\sum_{k=1}^{r}\sigma_{k}\ket{\lambda_k}\ket{u_k}\ket{v_k}
\xrightarrow{U_{\lambda,\tau}}\sum_{k=1}^{r}\sigma_{k}\ket{y_k}\ket{\lambda_k}\ket{u_k}\ket{v_k},
\end{align}
and its quantum circuit is given in Fig.~\ref{Fig U_lambdatau_}.
In addition, the registers stored $\ket{\tau}$, $\ket{y_k}$ respectively requires the same $n=O(log(\kappa))$ qubits. The number of the quantum gates required by the $U(\lambda,\tau)$ operation is $O(8\times (n+n))=O(16n)$.

\subsubsection{Unitary controlled operation \texorpdfstring{$CU\otimes I$}{Lg}} The purpose of this step is to employ unitary controlled operation \cite{divincenzo1998quantum} and ancillary qubit to tell whether or not the eigenvalue in the measured quantum bits corresponds to a principal component. If the $y_k>0(\lambda_k>\tau)$, unitary controlled operation will reverse the top qubit (ancillary qubit), otherwise it will do nothing. This procedure can be represented as:
\begin{align}
&\ket{0}\sum_{k=1}^{r}\sigma_{k}\ket{y_k}\ket{\lambda_k}\ket{u_k}\ket{v_k}\nonumber\\
\xrightarrow{CU}(&\ket{1}\sum_{k=1}^{t}\sigma_{k}\ket{y_k}\ket{\lambda_k}\ket{u_k}\ket{v_k}\nonumber\\
&+\ket{0}\sum_{k=t+1}^{r}\sigma_{k}\ket{0}\ket{\lambda_k}\ket{u_k}\ket{v_k}).
\end{align} 
In addition, the number of the quantum gates required by the $CU$ operation is $O(n)$.

\subsubsection{Unitary reverse operation \texorpdfstring{$I\otimes U^{\dagger}$}{Lg}} The purpose of this step is to remove the unnecessary registers that stored the $\ket{y_k}$ and $\ket{\lambda_k}$, we perform the reverse operation of $U_{\lambda,\tau}$ and $U_{PE}$. The operation procedure can be represented as:
\begin{align}
(&\ket{1}\sum_{k=1}^{t}\sigma_{k}\ket{y_k}\ket{\lambda_k}\ket{u_k}\ket{v_k}\nonumber\\
&+\ket{0}\sum_{k=t+1}^{r}\sigma_{k}\ket{0}\ket{\lambda_k}\ket{u_k}\ket{v_k})\nonumber\\
\xrightarrow{U^{\dagger}}(&\ket{1}\ket{0}\ket{0}\sum_{k=1}^{t}\sigma_{k}\ket{u_k}\ket{v_k}\nonumber\\
&+\ket{0}\ket{0}\ket{0}\sum_{k=t+1}^{r}\sigma_{k}\ket{u_k}\ket{v_k}).
\end{align}

In addition, the number of the quantum gates required by the $U^{\dagger}$ operation is $O(16n+n^2)$.

\subsubsection{Measurement} When we measure the qubits, if the top qubit (ancillary quibit) collapse to $1$, it implies that the state of remaining qubits is $\ket{\psi_{A_0}^{'}}=\sum_{k=1}^{t}\sigma_{k}\ket{u_k}\ket{v_k}$.

\subsubsection{The second phase estimation \texorpdfstring{$I\otimes U_{PE}$}{Lg}} 
\begin{figure}[H]
\centering
\includegraphics[scale=1.0]{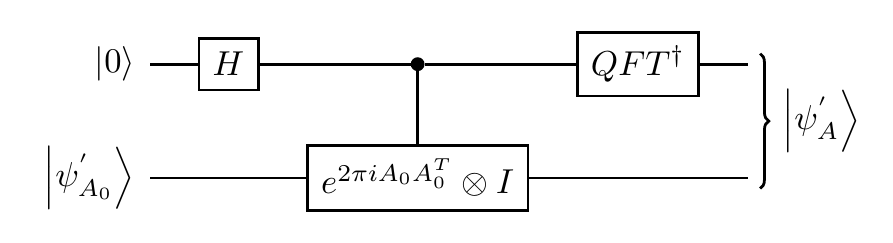}
\caption{\label{Fig our qPCA pe}The quantum circuit of the second phase estimation, where the input is $\ket{\psi_{A_0}^{'}}$ and the output is $\ket{\psi_{A}^{'}}$.}
\end{figure}

To obtain the quantum state $\ket{\psi_{A}^{'}}=\sum_{k=1}^{t}\sigma_{k}\ket{\lambda_k}\ket{u_k}\ket{v_k}$ from $\ket{\psi_{A_0}^{'}}=\sum_{k=1}^{t}\sigma_{k}\ket{u_k}\ket{v_k}$, we can perform another phase estimation on $\ket{\psi_{A_0}^{'}}$ shown in Fig.~\ref{Fig our qPCA pe}, i.e.
\begin{align}
\ket{\psi_{A_0}^{'}}\xrightarrow{U_{PE}(A)}\ket{\psi_{A}^{'}}.
\end{align}

In addition, the number of the quantum gates required by the second $U_{PE}$ operation is same $O(n^2)$.

\begin{figure}
\begin{algorithm}[H] 
\caption{\label{alg1}The low complexity qPCA algorithm.} 
 
\begin{flushleft}
\textbf{Input:}\\ 
A quantum state $\ket{\psi_{A_0}}$;\\
A unitary operation $U_{PE}(A)=e^{2\pi iA}$;\\
A threshold constant $\tau$.\\
\textbf{Output:}\\ 
A quantum state $\ket{\psi_{A}^{'}}$.\\
\textbf{Procedure:}\\
\end{flushleft}
\begin{algorithmic}[1]
\STATE Prepare quantum state\\
$\ket{\psi_1}=\ket{0}\ket{0}\ket{0}\ket{\psi_{A_0}}$.
\label{ code2:fram:prepare }
\STATE Perform the phase estimation $U_{PE}(A)$ to obtain\\
$\ket{\psi_2}=\ket{0}\ket{0}\sum_{1}^{r}\sigma_{k}\ket{\lambda_{k}}\ket{u_k}\ket{v_k}$. 
\label{code2:fram:pe}
\STATE Perform the unitary operation $U_{\lambda,\tau}$ to obtain\\
$\ket{\psi_3}=\ket{0}\sum_{k=1}^{t}\sigma_{k}\ket{y_{k}}\ket{\lambda_{k}}\ket{u_k}\ket{v_k}+\qquad\qquad\qquad\qquad$
$\ket{0}\sum_{k=t+1}^{r}\sigma_{k}\ket{0}\ket{\lambda_{k}}\ket{u_k}\ket{v_k}$. 
\label{code2:fram:trasform}
\STATE Perform the controlled operation $CU$ to obtain\\
$\ket{\psi_4}=\sum_{k=1}^{t}\sigma_{k}\ket{1}\ket{y_k}\ket{\lambda_{k}}\ket{u_k}\ket{v_k}+\qquad\qquad\qquad\qquad$
$\sum_{k=t+1}^{r}\sigma_{k}\ket{0}\ket{0}\ket{\lambda_{k}}\ket{u_k}\ket{v_k}$. 
\label{code2:fram:rotation}
\STATE Employ unitary operation $U_{\dagger}$ to obtain\\
$\ket{\psi_5}=\sum_{k=1}^{t}\sigma_{k}\ket{1}\ket{u_k}\ket{v_k}+\qquad\qquad\qquad\qquad$
$\sum_{k=t+1}^{r}\sigma_{k}\ket{0}\ket{u_k}\ket{v_k}$.
\label{code2:fram:reverse}
\STATE Measurement. When the measurement result of the top qubit is $1$, the quantum state will collapse to\\
$\ket{\psi_{A_0}^{'}}=\sum_{k=1}^{t}\sigma_{k}\ket{u_k}\ket{v_k}$. 
\label{code2:fram:measure}

\STATE Extract eigenvalues $\ket{\lambda_k}$ by performing the second phase estimation $U_{PE}$ to get\\ $\ket{\psi_A^{'}}=\sum_{k=1}^{t}\sigma_k\ket{\lambda_k}\ket{u_k}\ket{v_k}$. 
\label{code2:fram:phase}

\end{algorithmic}
\end{algorithm}
\end{figure}

The whole procedure of the proposed low complexity qPCA is shown in Algorithm~\ref{alg1} and the corresponding quantum circuit is shown in Fig.~\ref{Fig our qPCA}.

\section{Complexity and accuracy analysis}\label{sec3 analysis}
In this section, we analyze the circuit complexity and the accuracy of the proposed qPCA algorithm, and compared them with the state-of-the-art qPCA algorithm \cite{lin2019improved}.
\subsection{Circuit complexity}\label{sec3 a}
First, we analyze the number of qubits required for the  quantum circuit of our qPCA. As shown in Fig.~\ref{Fig our qPCA}, register Anc. contains only one ancillary qubit, and the number of the qubits in Reg. A, Reg. B and Reg. C are all $n=O(\log(\kappa))$. Finally, to save $\ket{\psi_{A_0}}$, it requires $O(\log(pq))$ qubits in Reg. M. Therefore the total qubits required for our qPCA is $O(\log(pq\kappa))$, which is the same as the qPCA in \cite{lin2019improved}.

Then we analyze the unitary operations of our qPCA algorithm contained in \eqref{our qPCA steps}, where each phase estimation $U_{PE}$ requires $O(n^2)$ quantum gates, and the operations $U(\lambda,\tau)$, $CU$ and $U^{\dagger}$  requires $O(16n)$, $O(n)$, and $O(n^2+16n)$ quantum gates, respectively. Therefore, the number of the quantum gates required by our qPCA algorithm is $O(3n^2+33n)$. As shown in Fig.~\ref{Fig ref qPCA circuit}, it requires much more quantum gates for the qPCA in \cite{lin2019improved}. The operations $U_{\sigma,\tau}$ and $U_{\sigma,\tau}^{'}$ both require $O(24n)$ quantum gates to compute the Newton's iterations. In addition, the three $U_{PE}$ operations, the two operations $R_y(\alpha)$, and the two operations $U^{\dagger}$ require $3O(n^2)$, $2O(n)$, and $2O(n^2+24n)$ quantum gates, respectively. Therefore the total number of the quanutm gates required for the qPCA in \cite{lin2019improved} is $O(5n^2+98n)$.

In summary, the proposed qPCA requires the same number of qubits and much less quantum gates compared with the previous qPCA in \cite{lin2019improved}. 

\subsection{The precision}\label{sec3 b}
The approximations of the qPCA in \cite{lin2019improved} take in two places: parameters estimation and Newton's iterations. However, the approximations of our qPCA only take in Newton's iterations, and even for the Newton's iterations, the proposed qPCA requires less quantum gates since its iterative function has lower order. Therefore, the proposed qPCA has higher level of precision.

\section{experiement}\label{sec4 experiement}
In this section, we perform experiments for our low complexity qPCA algorithm on the IBM quantum computing platform: IBM Quantum Experience \cite{cross2018ibm,balasubramanian10circuit,garcia2017five}.

\subsection{The experiment for the \texorpdfstring{$2\times2$}{Lg} matrix}\label{sec4 exp1}
Frist, we take the $2\times2$ matrix
\begin{equation}\label{Eq.17}
A={
\left[ \begin{array}{ccc}
1.5 & 0.5\\
0.5 & 1.5 
\end{array} 
\right ]}
\end{equation}
as an example, of which the quantum state is given by \cite{coles2018quantum} 
\begin{eqnarray}\label{Eq.19}
  \ket{\psi_A}=[0.6708,0.2236,0.2236,0.6708]^{T}. 
\end{eqnarray}
 Notice that the classical PCA should yield
\begin{align}
    &\lambda_1 = 2, \quad u_1 = [0.7071,0.7071]^{T},\nonumber \\
    &\lambda_2 = 1, \quad u_2 = [-0.7071,0.7071]^{T}.\nonumber
\end{align}

When we set the threshold $\tau=1$, only the eigenvectors $u_1$ with the eigenvalues $\lambda_1$ are reserved, the vector of the algorithm should be given by 
\begin{align}
\frac{\lambda_1 \ket{u_1}\ket{u_1}}{\sqrt{\lambda_{1}^{2}}}=[0.5000,0.5000,0.5000,0.5000]^{T}.
\end{align}

\begin{figure*}
	\centering
		\includegraphics[scale=.3]{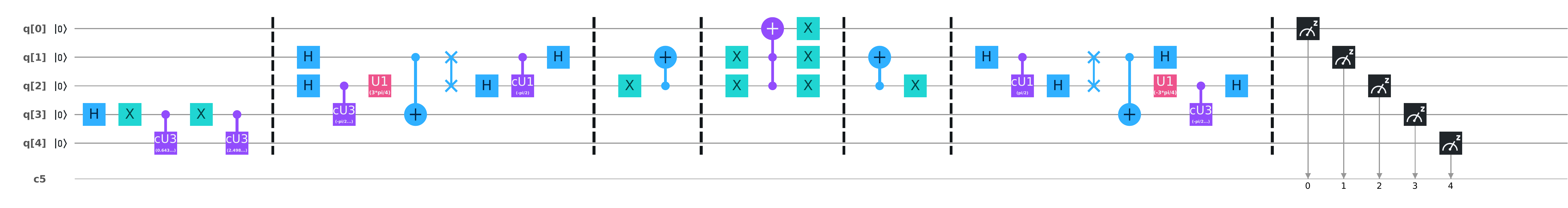}
    \caption
{\label{Fig circuit2}The experimental circuit of our qPCA for the $2\times2$ matrix $A$ with threshold $\tau=1$ on IBM Quantum Experience. The input of the quantum circuit is $\ket{\psi_A}$, and the output is $\ket{\psi_{A}^{'}}$. The qubit q[0] is an ancillary qubit. Before the first dash line of the quantum circuit, the qubits q[3-4]  are used to initialize the quantum state $\ket{\psi_A}$. Between the first dash and the second dash lines in the quantum circuit, the qubits q[1-2] are used to save eigenvalues from the phase estimation. Between the second and the third dash lines in the quantum circuit, the eigenvalues $\ket{\lambda_{k}}$ are converted to $\ket{y_{k}}$ on q[1-2]. Between the third and the fourth dash lines is the controlled operation. The rest of the quantum circuit are the inverse operations and the measurement.}

\end{figure*}
\begin{figure*}[ht]
		\includegraphics[scale=.097]{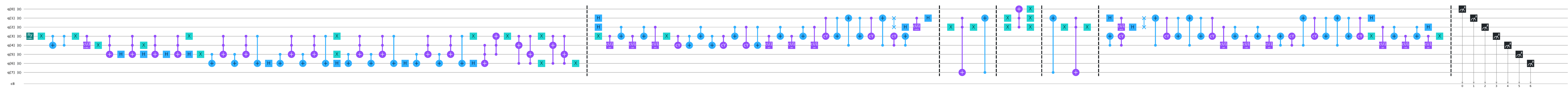}
    \caption
{\label{Fig circuit4}The experimental circuit of our qPCA for the $4\times4$ matrix $C$ with threshold $\tau=1.8$ on IBM Quantum Experience. The input is the quantum state $\ket{\psi_C}$, and the output is $\ket{\psi'_C}$. The qubits q[0] and q[7] are the ancillary qubits. Before the first dash line of the quantum circuit, the qubits q[3-6] are used to initialize the quantum state $\ket{\psi_{C}}$. Between the first and the second dash lines in the quantum circuit, the qubits q[1-2] in are used to save eigenvalues from the phase estimation. Between the second and the third dash lines in the quantum circuit, the eigenvalues $\ket{\lambda_{k}}$ is converted to $\ket{y_{k}}$ on q[1-2]. Between the third and the fourth dash lines is the controlled operation. The rest of the quantum circuit are the inverse operations and the measurement.}
\end{figure*}

The implementation of the quantum circuit for our qPCA algorithm on the IBM Quantum Experience is shown in Fig.~\ref{Fig circuit2}. Five qubits are required in total. The first qubit q[0] is used as an ancillary qubit. The second to third qubits q[1-2] are used to save eigenvalues $\ket{\lambda_k}$ and $\ket{y_k}$, and the qubits q[3-4] are used to initialize the quantum state $\ket{\psi_A}$. When the measurement result of q[0] is $1$, q[3-4] will collapse into the quantum state $\ket{\psi_{A}^{'}}$.

\begin{figure}
\centering
\subfigure[The probability histogram of our qPCA algorithm in the Circuit Composer from IBM Quantum Experience.]{
\begin{minipage}[b]{0.5\textwidth}\label{Fig results2(a)}
\includegraphics[scale=.6]{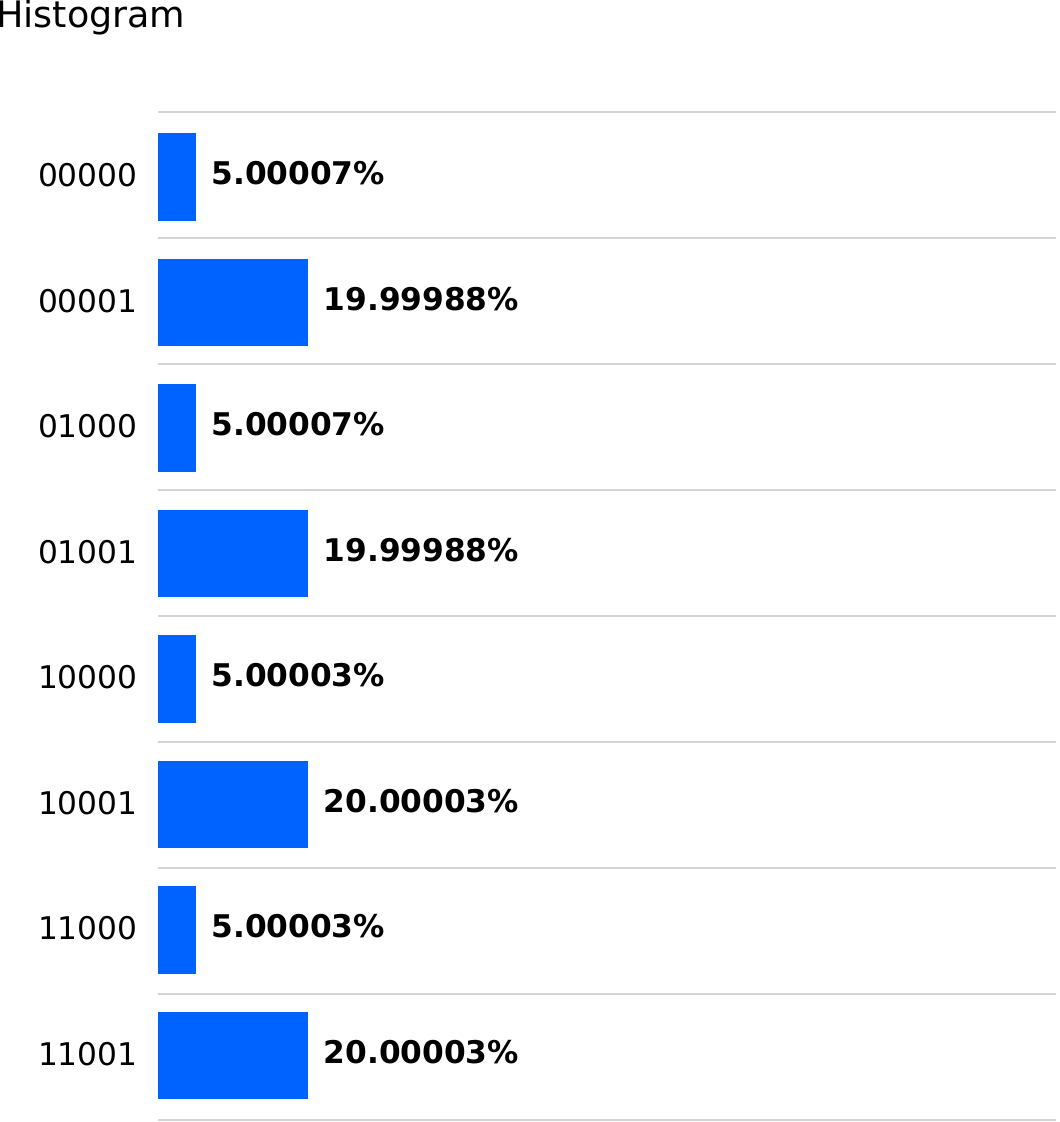}
\end{minipage}
}

\subfigure[The probability histogram of our qPCA algorithm in the QASM simulator from IBM Quantum Experience.]{
\begin{minipage}[b]{0.5\textwidth}\label{Fig results2(b)}
\includegraphics[scale=.21]{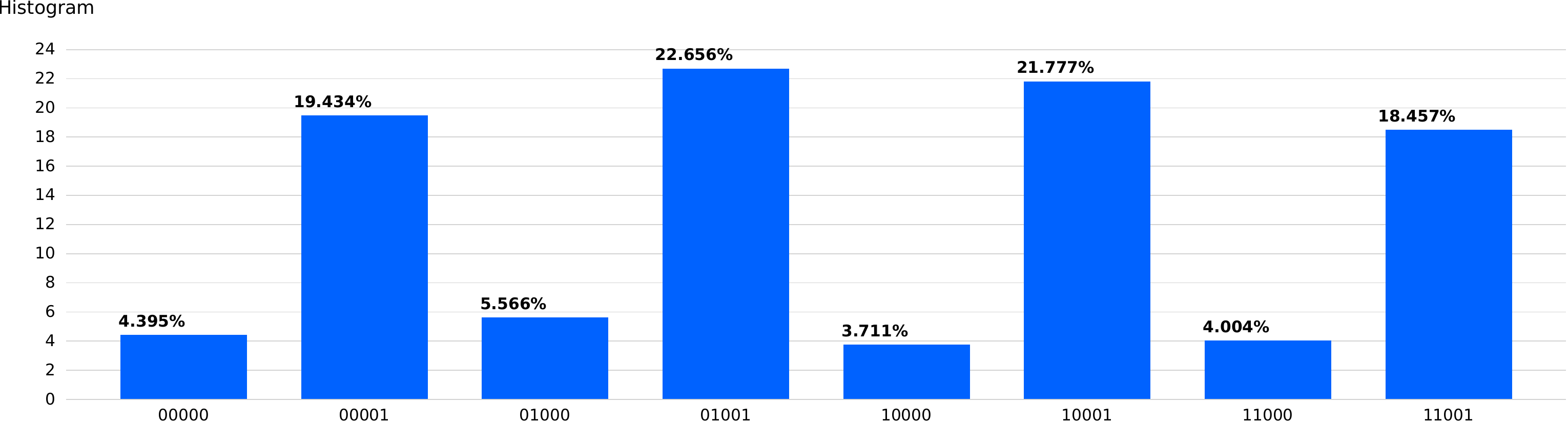}
\end{minipage}
}
\caption{\label{Fig results2}The Circuit Composer (theoretical) result and the QASM simulator result of our qPCA for the $2\times2$ matrix $A$ with threshold $\tau=1$ from IBM Quantum Experience.} 

\end{figure}

Circuit Composer on IBM Quantum Experience lets us see how quantum circuits affect the state of a collection of qubits through the measurement probabilities visualizations \cite{IBMQ-toolbox-web}. As shown in Fig.~\ref{Fig results2(a)}, when the top qubit measures $1$, the normalized vector of the statistical graph is given by
\begin{align}
\ket{\psi_A^{'}}_{circuit}=[0.5000,0.5000,0.5000,0.5000]^{T}.
\end{align}

The QASM simulator simulates the execution of quantum circuits and returns counts in histogram, then we run the quantum circuit on the the QASM simulator, and the result as shown in Fig.~\ref{Fig results2(b)} is given by
\begin{align}
\ket{\psi_A^{'}}_{qasm}=[0.4859,0.5245,0.5143,0.4735]^{T}.
\end{align}

Similarly, when we set the threshold $\tau=0.8$, the eigenvectors $u_1$, $u_2$ with corresponding the eigenvalues $\lambda_1$, $\lambda_2$ are reserved, the vector should be given by
\begin{align}
&\frac{\lambda_1 \ket{u_1}\ket{u_1} + \lambda_2 \ket{u_2}\ket{u_2}}{\sqrt{\lambda_1^{2}+\lambda_2^2}}\\
=&[0.6708,0.2236,0.2236,0.6708]^{T}.  
\end{align}and the result of the Circuit Composer is given by
\begin{align}
\ket{\psi_A^{'}}_{circuit}=[0.6708,0.2236,0.2236,0.6708]^{T},
\end{align}
the result of QASM simulator is given by
\begin{align}
\ket{\psi_A^{'}}_{qasm}=[0.6651,0.2296,0.2119,0.6782]^{T}.
\end{align}

\begin{figure}
\centering
\subfigure[The probability histogram of our qPCA algorithm in the Circuit Composer from IBM Quantum Experience.]{
\begin{minipage}[b]{0.5\textwidth}\label{Fig results4(a)}
\includegraphics[scale=.55]{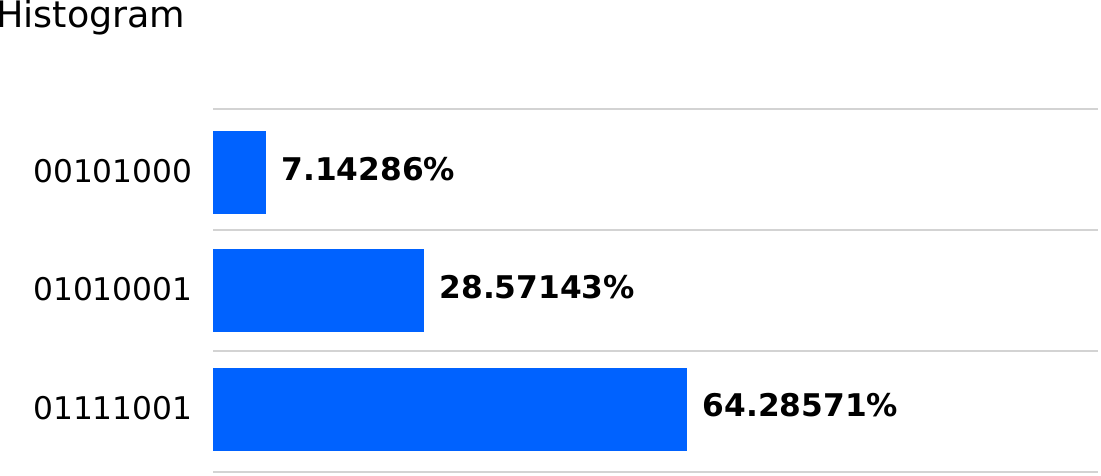}
\end{minipage}
}

\subfigure[The probability histogram of our qPCA algorithm in the QASM simulator from IBM Quantum Experience.]{
\begin{minipage}[b]{0.5\textwidth}\label{Fig results4(b)}
\includegraphics[scale=.25]{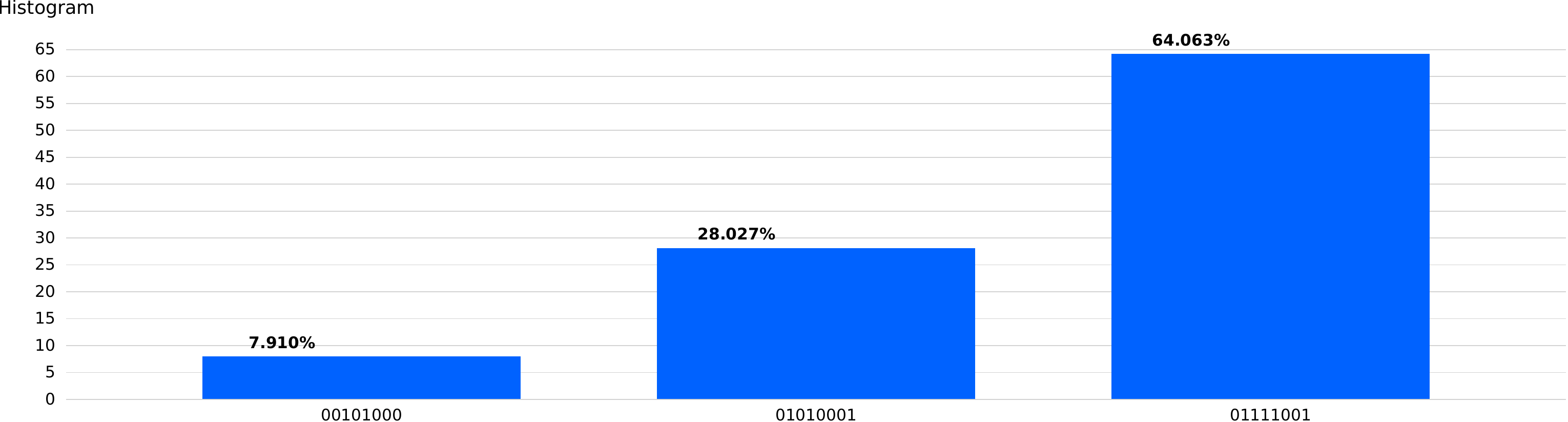}
\end{minipage}
}
\caption{ \label{Fig results4}The Circuit Composer (theoretical) result and the QASM simulator result of our qPCA for the $4\times4$ matrix $C$ with threshold $\tau=1.8$ from IBM Quantum Experience.}
\end{figure}

\subsection{The experiment for the \texorpdfstring{$4\times4$}{Lg} matrix}\label{sec4 exp2}
Now we take the $4\times4$ matrix
\begin{equation}
C={
\left[ \begin{array}{ccccccc}
0 & 0 & 0 & 0\\
0 & 1 & 0 & 0\\
0 & 0 & 2 & 0\\
0 & 0 & 0 & 3
\end{array} 
\right ]}
\end{equation}
as another example, of which the corresponding quantum state is given by
\begin{align}\label{Eq.28}
\ket{\psi_{C}}=[\dots, 0.2673, \dots, 0.5345, \dots, 0.8018]^{T},
\end{align}
where the values $0.2673$, $0.5345$, $0.8018$ respectively represent the 6th, 11th, and 16th elements of the vector $\ket{\psi_C}$, and the rest are $0$.

Notice that the classical PCA should yield
\begin{align}
&\lambda_1 = 0,\quad u_1 = [1,0,0,0]^{T},\nonumber \\
&\lambda_2 = 1,\quad u_2 = [0,1,0,0]^{T},\nonumber \\
&\lambda_3 = 2,\quad u_2 = [0,0,1,0]^{T},\nonumber \\
&\lambda_4 = 3,\quad u_2 = [0,0,0,1]^{T}.\nonumber
\end{align}
When we set the threshold $\tau=1.8$, only the eigenvectors $u_3$, $u_4$ with corresponding the eigenvalues $\lambda_3$, $\lambda_4$ are reserved, the vector should be given by:
\begin{align}
&\frac{\lambda_3 \ket{u_3}\ket{u_3} + \lambda_4 \ket{u_4}\ket{u_4}}{\sqrt{\lambda_3^{2}+\lambda_4^2}}\nonumber\\
=&[\dots,0.0000,\dots, 0.5547,\dots, 0.8321]^{T},
\end{align}
where the values $0.0000$, $0.5547$, $0.8321$ respectively represent the 6th, 11th, and 16th elements of the vector, and the rest are $0$.

The implementation of the quantum circuit for our low complexity qPCA algorithm for the matrix $C$  with $\tau=1.8$ is shown in Fig.~\ref{Fig circuit4}. The qubits q[0] and q[7] are ancillary qubits and the qubits q[1-2] stored the eigenvalues. The qubits q[3-6] are used to prepare the initial quantum state $\ket{\psi_{C}}$. When the measurement result of q[0] is $1$, q[3-6] will collapse into the quantum state $\ket{\psi_C^{'}}$. The initial quantum state $\ket{\psi_C}$ and the phase estimation operation in the implementation are not straightforward to construct. Therefore we design the binary tree to prepare the initial quantum state, as shown in Fig.~\ref{bianry} of Appendix~\ref{sec app1}, and the corresponding quantum circuit is shown in Fig.~\ref{binary circuit}. The quantum circuit of the phase estimation on $\ket{\psi_{C}}$ is shown in Fig.~\ref{c-u decomposition} of Appendix~\ref{sec app2}.

As shown in Fig.~\ref{Fig results4}, the result of the Circuit  Composer is given by
\begin{eqnarray}
\ket{\psi_C^{'}}_{circuit}=[\dots,0.0000,\dots, 0.5547, \dots, 0.8321]^{T},
\end{eqnarray}
and the result of the QASM simulator is given by
\begin{eqnarray}
\ket{\psi_C^{'}}_{qasm}=[\dots,0.0000,\dots, 0.5294,\dots,0.8004]^{T},
\end{eqnarray}
where the values $0.0000$, $0.5547$, $0.8321$ respectively represent the 6th, 11th, and 16th elements of the vector $\ket{\psi_C^{'}}_{circuit}$, the values $0.0000$, $0.5294$, $0.8004$ respectively represent the 6th, 11th, and 16th elements of the vector $\ket{\psi_C^{'}}_{qasm}$, and the rest are $0$.

Similarly, when we set the threshold $\tau=0.5$, the eigenvectors $u_2$, $u_3$, $u_4$ with the corresponding eigenvalues $\lambda_2$, $\lambda_3$, $\lambda_4$ are reserved, the vector should be given by:
\begin{align}
&\frac{\lambda_2\ket{u_2}\ket{u_2}+\lambda_3\ket{u_3}\ket{u_3}+\lambda_2\ket{u_2}\ket{u_2}}{\sqrt{\lambda_2^{2}+\lambda_3^{2}+\lambda_4^{2}}}\nonumber \\
=&[\dots,0.2673,\dots,0.5345,\dots, 0.8018]^{T},
\end{align}
where the values $0.2673$, $0.5345$, $0.8018$ respectively represent the 6th, 11th, and 16th elements of the vector and the rest are $0$.
The result of the Circuit Composer is given by\begin{align}
\ket{\psi_C^{'}}_{circuit}=[\dots,0.2673,\dots, 0.5345,\dots, 0.8018]^{T},
\end{align}
and the result of the QASM simulator is given by
\begin{align}
\ket{\psi_C^{'}}_{qasm}=[\dots,0.2500,\dots,0.5484,\dots, 0.7979]^{T},
\end{align}
where the values $0.2673$, $0.5345$, $0.8018$ respectively represent the 6th, 11th, and 16th elements of the vector $\ket{\psi_C^{'}}_{circuit}$, the values $0.2500$, $0.5484$, $0.7979$ respectively represent the 6th, 11th, and 16th elements of the vector $\ket{\psi_C^{'}}_{qasm}$, and the rest values are $0$.

Based on the the experimental results of $2\times2$ and $4\times4$ matrices with different threshold respectively, we can see that the Circuit Composer (theoretical) results yielded by our qPCA algorithm are exactly the same as that of classical PCA. For the results yield by quantum computer simulator, our algorithm can also obtain high accuracy. In short, the experimental results meet our expectations.

\section{Conclusion}\label{sec5 conclusion}
In this paper, we propose a low complexity qPCA algorithm, which outputs the quantum state only containing the principal components. This is similar to the state-of-the-art qPCA \cite{lin2019improved} algorithm. The advantages of the proposed qPCA are as follows: The number of quantum gates required is only about $3/5$ of that of the state-of-the-art; we also show that it has a higher level of precision. Finally we implement the proposed qPCA on IBM Quantum Experience, and the experimental results support our expectations. 

\appendices
\subsection{The preparation of state \texorpdfstring{$\ket{\psi_C}$}{Lg}}\label{sec app1}
\begin{figure}
	\centering
		\includegraphics[scale=.56]{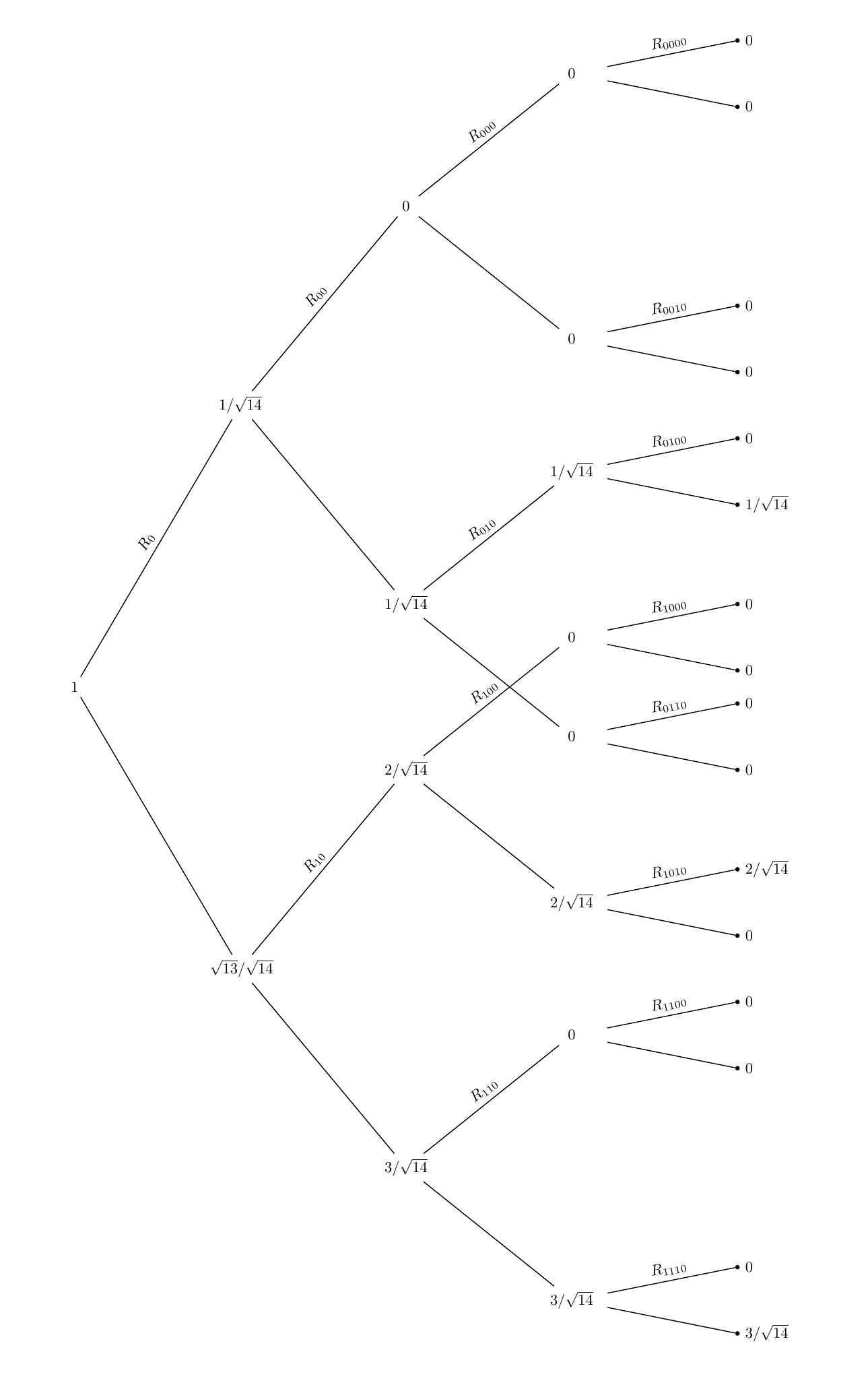}
	\caption{\label{bianry}The binary tree to prepare the state $\ket{\psi_C}$.}
\end{figure}

The initial state $\ket{\psi_C}$ in \eqref{Eq.28} is not straightforward to prepare on IBM Quantum Experience. Therefore we design the binary tree \cite{grover2002creating,kerenidis2016quantum} as shown in Fig.~\ref{bianry} to prepare the quantum state, whose leaf nodes are the vectors of the quantum state, and each branch is a $R_y(\theta)$ unitary operation, where
 \begin{equation}
R_y(\theta)={
\left[ \begin{array}{ccc}
\cos(\frac{\theta}{2}) & -\sin(\frac{\theta}{2})\\
\sin(\frac{\theta}{2}) & \cos(\frac{\theta}{2}) 
\end{array} 
\right ]}.
\end{equation}

\begin{figure*} 
	\centering
		\includegraphics[scale=1.0]{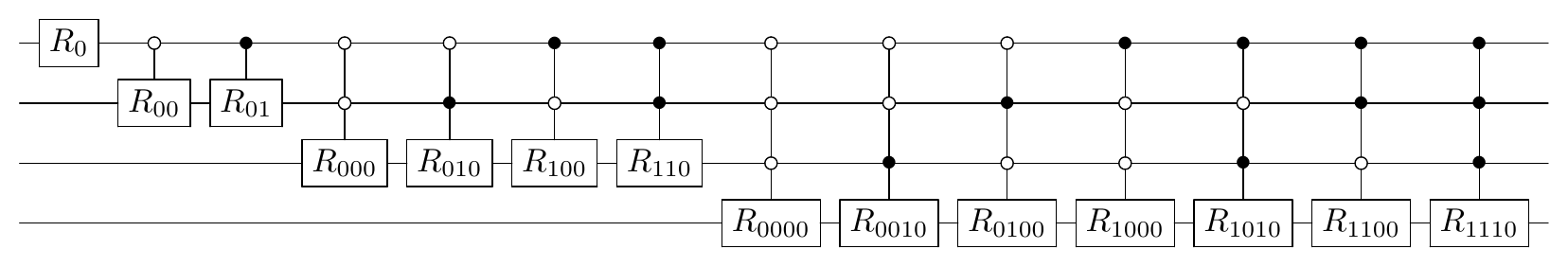}
\caption{\label{binary circuit}The quantum circuit to prepare the state $\ket{\psi_C}$.}
\end{figure*}

The corresponding quantum circuit of the binary tree is shown in Fig.~\ref{binary circuit}.

\subsection{The phase estimation of matrix \texorpdfstring{$C$}{Lg}}\label{sec app2}
\begin{figure}
\subfigure[Unitary decomposition of $C-U$.]{
\begin{minipage}[b]{0.5\textwidth}
\centering
\includegraphics[width=.43\linewidth,height=1.8cm]{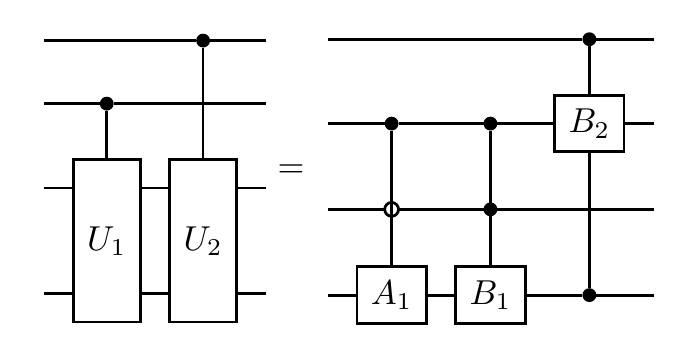}\label{Fig10(a)}
\end{minipage}
}

\subfigure[Unitary decomposition of $A_1$.]{
\begin{minipage}[b]{0.5\textwidth}
\centering
\includegraphics[width=.77\linewidth]{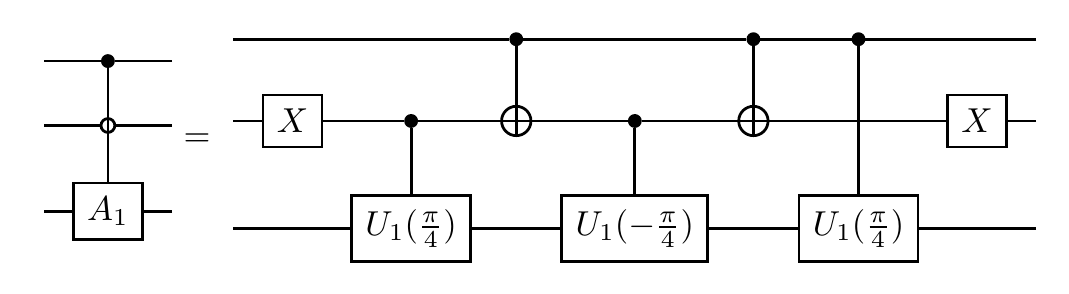}\label{Fig10(b)}
\end{minipage}
}

\subfigure[Unitary decomposition of $B_1$.]{
\begin{minipage}[b]{0.5\textwidth}
\centering
\includegraphics[width=.83\linewidth]{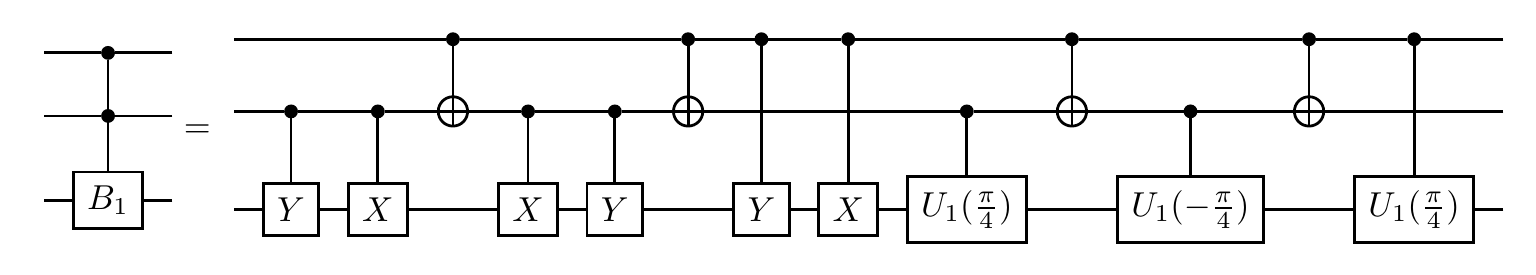}\label{Fig10(c)}
\end{minipage}
}

\subfigure[Unitary decomposition of $B_2$.]{
\begin{minipage}[b]{0.5\textwidth}
\centering
\includegraphics[width=.67\linewidth]{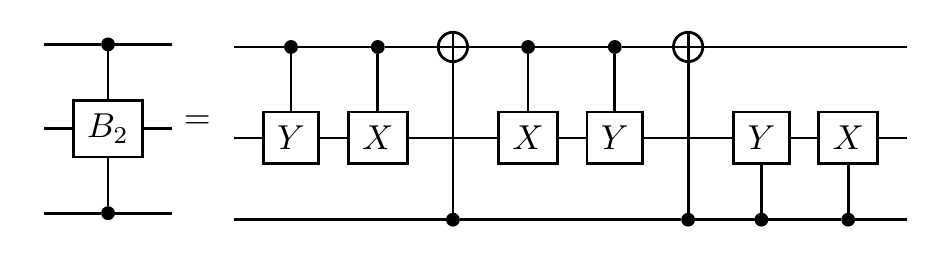}\label{Fig10(d)}
\end{minipage}
}
\caption{\label{c-u decomposition}The unitary operation of the phase estimation of the matrix $C$.}

\end{figure}

The quantum circuit of the phase estimation on the matrix $C$ is not straightforward to design on IBM Quantum Experience. Therefore we decompose the phase estimation into several unitary operations which can be implemented by simple quantum gates \cite{vartiainen2004efficient,li2013decomposition}. The unitary matrices in the phase estimation of $C$ are $U_1=e^{\frac{2\pi{iC}}{4}}$,  $U_2=e^{\frac{2\pi{iC}}{2}}$ \cite{mohammadbagherpoor2019experimental,dutta2018demonstration}, where 
\begin{equation}
\begin{matrix}
U_1=\begin{bmatrix}
0 & 0 & 0 &  0 \\ 
0 & i  &0 &  0 \\
0 & 0 & -1 &  0 \\
0 & 0 & 0 &  -i 
\end{bmatrix} ,& U_2= \begin{bmatrix}
0 & 0 & 0 & 0\\
0 & -1 & 0 & 0\\
0 & 0 & 1 & 0\\
0 & 0 & 0 & -1
\end{bmatrix}
\end{matrix}.
\end{equation}
The corresponding quantum circuits of $C-U_{1}$, $C-U_{2}$ are show in Fig.~\ref{c-u decomposition}.
\bibliographystyle{IEEEtran}
\bibliography{IEEEabrv,ref}
\EOD
\end{document}